\documentstyle[preprint,prb,aps]{revtex}
\begin{document}

\draft

\title{Structural and insulator-metal quantum phase transitions on a
lattice}
\author{E. V. Tsiper, A. L. Efros}
\address{Department of Physics, University of Utah, Salt Lake City,
Utah 84112}

\date{\today}
\maketitle

\begin{abstract}
We consider 2D gas of spinless fermions with the Coulomb and the short
range interactions on a square lattice at $T=0$.  Using exact
diagonalization technique we study finite clusters up to 16 particles
at filling factors $\nu=1/2$ and 1/6.  By increasing the hopping
amplitude we obtain the low-energy spectrum of the system in a wide
range from the classical Wigner crystal to almost free gas of
fermions.  The most efforts are made to study the mechanism of the
structural and insulator-metal transitions.  We show that both
transitions are determined by the energy band of the defect with the
lowest energy in the Wigner crystal.
\end{abstract}

\section{Introduction}

The insulator-metal (IM) transition and the role of electron-electron
interaction in this transition is a problem of permanent interest,
both theoretical and experimental.  It has been
shown\cite{pik1,ber,shep} that in the systems with strong disorder the
interaction is in favor of delocalization because electrons may help
each other to overcome the random potential.  In clean systems the
role of the interaction is opposite.  It may create the so-called
correlated insulator in a system which would be metallic otherwise.
The Wigner crystal (WC) is a good example of such insulator.

WC in continuum is not an insulator itself, since it can move as the
whole and carry current.  However, due to shear modulus it can be
pinned by a small disorder.  The ground-state energy of the continuum
WC and its zero-temperature melting was widely studied in the recent
years both with and without magnetic field.\cite{zhu}

In contrast to the continuum case, the WC on a lattice can be an
insulator without any disorder due to the Umklapp processes in a host
lattice.  The WC on a lattice does not have any sound or soft plasma
modes and its excitation spectrum has a gap.

The great majority of the efforts made recently to study correlated
particles on a lattice were restricted to the Hubbard model or $t-J$
model (See review Ref.~\onlinecite{dag}).  The so-called extended
Hubbard model with short-range and long-range interactions has been
mostly considered for bosons in connection to the
insulator-superconductor transition.\cite{zim,ann,er,er1} In these
papers supersolid and superfluid phases have been found.  The bosons
with infinite on-site repulsion are called hard-core bosons.  In the
case of the nearest-neighbor interaction the hard-core boson problem
maps into Ising-Heisenberg spin Hamiltonian.

Spinless fermions are similar to the hard-core bosons.  In both
systems the number of particles on a site is either zero or one. In
1D-case these two systems are equivalent if the interaction between
particles does not permit them to penetrate through one
another.\cite{Girardeau}

The 1D problem with the nearest-neighbor interaction at half filling
is exactly soluble.\cite{yang,des,suth} This instructive solution shows
that the transition is not of the first order and that the
IM transition, as detected by the stiffness constant,
appears at the same point as the structural transition.\cite{suth}

Very few works exist on the extended Hubbard model for 2D fermions.
Pikus and Efros\cite{pik} have performed a computer modeling for 2D
spinless fermions with Coulomb interaction on a square lattice at
filling factors $\nu=1/3$ and $1/6$.  They argue that the lifting of
the ground-state degeneracy with increasing $J$ is a very good
diagnostic of the structural phase transition.  They have also found a
similarity between the systems of spinless fermions and hard-core
bosons near the transition.

In this paper we study structural and IM transitions for spinless
fermions at $\nu=1/2$ and 1/6.  To detect these transitions we use the
ground-state splitting and the flux sensitivity\cite{kohn,scal}
respectively.  The purpose of the work is to take advantage of the
exact diagonalization technique and to study the modification of the
low-energy part of the spectrum in a wide interval of the hopping
amplitude $J$ all the way from the classical WC to the free fermion
limit.

Our results for long-range and short-range interactions suggest a
simple picture of the transition.  The transition is related mainly to
the modification of the two lowest branches of energy spectrum.  At
small $J$ these two branches are the WC state and the energy band of
the defect with the lowest energy.

The paper is organized as follows.  In Sec.~II we describe our
numerical technique and present some general results.  Sec.~III
contains the results of computations and their discussion.  We suggest
the mechanism of the transition, analyze the role of the size effect
in finite-cluster computations, and discuss the possibility that the
delocalized phase above the IM transition is superconducting.  The
better to illustrate the mechanism of the transition we present the
dependence of the total energy on the quasimomentum, $E(P)$, for a 1D
system with $1/r$ interaction.

\section{Computational approach and general remarks}

We consider spinless fermions at $T=0$ on the 2D square lattice
described by the following model Hamiltonian
\begin{equation}
H=J\sum_{{\bf r},{\bf s}}a_{{\bf r}+{\bf s}}^\dagger a_{\bf r}
  \exp(i\bbox{\phi s})
+ \frac{1}{2}\sum_{{\bf r}\neq{\bf r}^{'}}
 n_{\bf r}n_{{\bf r}{'}}V(|{\bf r}-{\bf r}{'}|).
\label{ham}
\end{equation}

Here $n_{\bf r}=a_{\bf r}^\dagger a_{\bf r}$, the summation is
performed over the lattice sites {\bf r}, {\bf r}$^{'}$ and over the
vectors of translations {\bf s} to the nearest-neighbor sites.  We
consider long-range (LR) Coulomb potential $V(r)=1/r$ and short-range
(SR) strongly screened Coulomb potential $V(r)=\exp(-r/r_s)/r$ with
$r_s=0.25$ in the units of lattice constant.  We study rectangular
clusters $L_x\times L_y$ with the periodic boundary conditions.  The
dimensionless vector potential $\bbox{\phi}=(\phi_x,\phi_y)$ in the
Hamiltonian is equivalent to the twist of the boundary conditions by
the flux $\Phi_i=L_i\phi_i$, $i=x,y$.  The energy spectrum is periodic
in $\Phi_x$ and $\Phi_y$ with the period $2\pi$.

As a basis for computations we use many-electron wave functions at
$J=0$ in the coordinate representation: $\Psi_\alpha=\prod_{i=1}^N
a_{{\bf r}_i}^\dagger|{\rm VAC}>$.  The total size of the Hilbert
space is $C_M^N$, where $M=L_x\times L_y$ is the area of a system, and
$N$ is the number of particles.

The basic functions $\Psi_\alpha$ can be visualized as pictures, which
we call {\em icons}.  Some lowest energy icons are shown in
Fig.~\ref{fig1}.  The energy of each icon is calculated as a Madelung
sum, assuming that the icon is repeated periodically over the infinite
plane with a compensating homogeneous background.

The icon with the lowest energy is a fragment of the crystal.  The
icons with higher energies represent different types of defects in
WC.

\subsection{Quasimomentum representation}

The Hamiltonian Eq.~(\ref{ham}) is translationally invariant.  For
each icon $\alpha$ there are $m_\alpha$ different icons that can be
obtained from it by various translations.  These icons are combined to
get the wave function with total quasimomentum {\bf P}:
\begin{equation}
  \Psi_{\alpha{\bf P}}=\frac{1}{\sqrt{m_\alpha}}
  \sum_{\bf r}\exp(i{\bf P}{\bf r})T_{\bf r}\Psi_\alpha.
\label{func}
\end{equation}
The summation is performed over $m_\alpha$ translations $T_{\bf r}$.
This transformation reduces the effective Hilbert space size by
approximately $M$ times.

For the icons with periodic structures the number $m_\alpha$ of
different functions $\Psi_{\alpha{\bf P}}$ is smaller than $M$.  For
example, the icon $\Psi_0$ of the WC with one electron per primitive
cell generates $m_0=1/\nu$ different values of {\bf P}.  These values
are determined by the conditions
\begin{equation}
(-1)^{Q_j}\exp(i{\bf P}\bbox{l}_j)=1.
\end{equation}

Here $\bbox{l}_j$ are the primitive vectors of the WC, and $Q_j$ are
the numbers of fermionic permutations necessary for translations on
these vectors.  These conditions can be easily understood.  If
translation on a vector $\bbox{l}_j$ is applied to Eq.~(\ref{func}),
the right-hand side acquires a factor $(-1)^{Q_j}$, while for a
function with given {\bf P} this factor must be equal to $\exp(i{\bf
P}\bbox{l}_j)$.  If $Q_j$ are even for both $\bbox{l}_j$, the allowed
{\bf P} form the reciprocal lattice of the WC.  However, in the case
when one or both of $Q_j$ are odd, the lattice is shifted by $\pi$ in
the corresponding directions.  In such case ${\bf P}=0$ is forbidden.
The complete set of $m_\alpha$ nontrivial values of {\bf P} can be
obtained by restricting {\bf P} to the first Brillouin zone of the
background lattice.  One WC is represented by a number of icons
obtained from each other by the point-group transformations of the
background lattice.

Note that the total number of allowed values of {\bf P} for the WC is
the property of the WC and it remains finite at infinite cluster size.
Contrary, an icon representing a point defect in a WC generates all
vectors {\bf P}.  Their total number is equal to the volume $M$ of the
first Brillouin zone of the background lattice.

\subsection{General remarks}

In the macroscopic system all the states generated by the WC icon form
the ground state degenerate at small $J$.  This degeneracy appears
because the effective matrix elements which connect translated WC's
are zero in the macroscopic limit.  The total energy as a function of
quasimomentum {\bf P} has identical minima at all {\bf P} generated by
the WC icons.  The spectra of excitations in the vicinity of these
minima are also identical.

The charge density for the state with given quasimomentum {\bf P} (see
Eq.~\ref{func}) is always the same at all sites of the host lattice.
However, at small $J$ the correlation function indicates a long range
order.  Any small perturbation, which violates translational
invariance, splits the degeneracy in such a way that the ground state
describes a single WC with a strong modulation of the charge density.

The lifting of the ground state degeneracy at some critical value
$J_c$ indicates a structural phase transition and restoration of the
host lattice symmetry.

The flux sensitivity of a macroscopic system is zero at small $J$.  It
becomes non-zero at some finite value of $J$ which might be different
from $J_c$.  We associate this transition with the IM
transition\cite{kohn}.

For the finite system the following results can be obtained directly
using the perturbation theory with respect to $J$:

(i) the ground state and the lowest excited states have a large common
negative shift which is proportional to $J^2$ and to the total number
of particles $N$. This shift is the same for all low-lying states and
does not affect the excitation spectrum of the system;

(ii) at $\nu=1/2$ the splitting of the ground state appears in the
$N$-th order and is proportional to $J^N$. At other filling factors
the degeneracy of the ground state at $J=0$ is larger than two.  The
splitting is determined by matrix elements which are proportional to
$J^K$.  For each matrix element the value of $K$ is equal to the
number of hops necessary to obtain one crystalline structure from
another and is proportional to $N$;

(iii) the flux dependence of the ground state for the flux in
$x$-direction appears in the $L_x$-th order and is proportional to
$J^{L_x}$ in 2D case.  In 1D the flux dependence appears in the $N$-th
order and is proportional to $J^N$.

Thus, we conclude that both lifting of the ground-state degeneracy and
appearance of the flux sensitivity occur very sharply and they can be
used as convenient criteria for the structural and the IM transitions
respectively.  Note that the correlation function is a less sensitive
criterion for small clusters\cite{pik,old} since it does not exhibit
sharp behavior in the transition region.

\section{Results of computations and discussion.}

\subsection{Results of Computations.}

Fig.~\ref{fig2}a,b shows the results of diagonalization for cluster
$4\times6$ with 12 electrons for the LR (a) and SR (b) interactions.
The total energy $E$ is shown as a function of $J$.  The ground state
energy is taken as a reference point for $E$.  Here and below, the
unit of energy is the LR interaction energy between nearest neighbors.
At $J=0$ the values of $E$ coincide with the energies of the icons
shown in Fig.~\ref{fig1}.  We define $\Delta$ as the gap between the
ground and first excited states at $J=0$.  Note that $\Delta$ in the
LR case is almost exactly 10 times larger than in the SR case (see
Fig.~\ref{fig1}a,b).

At large $J$ the energy $E$ is linear in $J$.  Thus, we can conclude
that with increasing $J$ in this interval we go all the way from
classical icons to free fermions.  The ground state is almost
degenerate at small $J$ and it splits into two states with increasing
$J$.  As we have discussed above, this is a manifestation of the
structural transition.  The quasimomenta of these two states, ${\bf
P}=(0,\pi)$ and $(\pi,0)$, are those generated by the WC icon.  In
Fig.~\ref{fig2}a,b they are denoted as (0,3) and (2,0), where
$(n_x,n_y)$ stands for quasimomentum with projections $P_x=2\pi
n_x/L_x$, $P_y=2\pi n_y/L_y$.  The other branches are the bands of
defects.

Fig.~\ref{fig3}a,b shows flux sensitivity $\delta E=|E(\pi)-E(0)|$,
computed for the ground state for two directions of the vector
potential.  Here $E(\Phi)$ stands for the total energy as a function
of $\Phi_x$ or $\Phi_y$.  In accordance with perturbation theory (see
Sec.~II B), the flux sensitivity at small $J$ obeys the laws $J^4$ and
$J^6$ for the direction of the vector potential along the short and
long sides of the cluster respectively.  The energy splitting between
the lowest states with ${\bf P}=(0,\pi)$ and $(\pi,0)$ is also shown.
At small $J$ the splitting is proportional to $J^{12}$ (12 is the
number of particles), as it follows from the perturbation theory.

At large $J$ the flux sensitivity is linear in $J$ and coincides with
the free-fermion value.  Note that for free fermions at $\nu=1/2$ the
flux sensitivity $\delta E$ is size independent for large
clusters.\cite{free}

The intervals $\delta J$ where computational curves for $\delta E$
make a crossover from one asymptotic to another are pretty narrow.  In
what follows we assume that these are the critical intervals for the
IM transition, smeared in a finite cluster.  These critical intervals
can be fairly well defined for each cluster and should shrink into a
transition point with increasing cluster size.

The vertical bars in Fig.~\ref{fig3}a,b show the estimated critical
interval $\delta J$ for the IM transition which is approximately
0.15---0.25 for the LR interaction and 0.015---0.025 for the SR
interaction in a cluster $4\times6$.

The behavior of the ground state splitting (GSS) at large $J$ is more
complicated.  In the free-fermion approximation the GSS is zero.
Considering interaction as perturbation one can show that in a
$4\times 4$-cluster the GSS $\rightarrow0$ as $J\rightarrow\infty$.
In a $4\times 6$-cluster the GSS $\rightarrow 5.76\times 10^{-5}$ for
the SR interaction and the GSS $\rightarrow 0.14$ for the LR
interaction as $J\rightarrow\infty$.  These analytical calculations
are in a good agreement with the computational results at large $J$
given in Fig.~\ref{fig3}a,b.  In the case of SR interaction the GSS
curve has maximum.  It is reasonable to assume that the crossover from
WC to free fermions occurs in the vicinity of this maximum.  There is
no maximum in the case of LR interaction and the crossover region can
be estimated using the sharp maximum of the second derivative.

We conclude that within accuracy of our computations, limited by the
finite cluster sizes, the IM transition detected by the flux
sensitivity and structural transition detected by the GSS occur
simultaneously.

Comparison of Fig.~\ref{fig3}a and \ref{fig3}b shows that the
dependencies $\delta E(J)$ for the LR and SR potentials are almost
indistinguishable if all the energy scales for one of them are
adjusted 10 times.  This factor is just the ratio of zero-$J$ gaps
$\Delta$ for these two cases.  Thus, we come to a conclusion that
$J_c$ depends on the type of interaction potential mostly through the
value of $\Delta$.  The same applies to the general structure of the
low-energy spectrum of the system in the transition region as can be
seen from comparison of Fig.~\ref{fig2}a and \ref{fig2}b.

Fig.~\ref{fig4}a,b shows the data for $\nu=1/6$ and LR interaction.
Fig.~\ref{fig4}a looks more complicated than Fig.~\ref{fig2}a,b.  The
WC for $\nu=1/6$ is shown in the first icon in Fig.~\ref{fig1}c.
There are four such WCs which can be obtained from each other by
point-symmetry operations.  Each WC generates six different values of
{\bf P}.  Thus, at small $J$ the ground state of the system is 24-fold
degenerate.  The degeneracy is high, however it remains the same in
the infinitely large cluster.

The primitive vectors of the WC at $\nu=1/6$ can not be obtained from
each other by any symmetry operation on the host lattice.  This means
that the WC phase belongs to a {\em reducible} representation of the
symmetry group of the host lattice.  Following Landau and
Lifshitz\cite{land}, the symmetry reduction in the second-order phase
transition should be such that the low-symmetry phase belongs to an
irreducible representation of the symmetry group of the high-symmetry
phase.  We conclude that the single second order phase transition is
forbidden in this case.  However, it can occur as a series of
transitions, each reducing the symmetry one step further.  In fact,
Fig.~\ref{fig4}a reminds the picture of multiple transitions.  We
think that each splitting of the energy levels generated by the WC
icon manifests a structural transition.  The cluster $6\times6$ is too
small to distinguish the critical intervals for each of these
transitions.  We can only conclude from Fig.~\ref{fig4}a,b that the
critical interval $\delta J$ for all of the structural transitions and
IM transition is 0.01---0.03. This interval is shown by vertical bars
in Fig.~4b.

\subsection{The mechanism of transition}

Our data suggest the following mechanism of the transition.  The width
of the band of the lowest defect in the WC increases with $J$ such
that its lowest edge comes close to the energy of the ground
state\cite{old} (see Figs.~\ref{fig2}a,b, \ref{fig4}a, and
\ref{fig5}).  Strong mixing between the crystalline and defect states
with the same quasimomentum occurs at this point.  The avoided
crossing appears between the ground state and the states in the defect
band.

One can interpret the avoided crossing in terms of the ground state
which acquires a large admixture of defect states.  This
interpretation reminds the idea of zero-point defectons proposed by
Andreev and Lifshitz.\cite{AL}

In principle, one can imagine that the state with a quasimomentum {\bf
P} different from those generated by the WC icon becomes the ground
state via a branch crossing.  However, in all cases we have
considered, we observe the avoided crossing between the crystalline
state and the state in the defect band with the same {\bf P}.
Assuming that this is the case for larger clusters, we conclude that
the phase transition is not of the first order.

The proposed mechanism of the transition can be illustrated by the
dependence $E({\bf P})$ at given $J$.  Unfortunately, in 2D case the
number of discreet values of {\bf P} along any line in the first
Brillouin zone is small even for the largest 2D system we study.  To
clarify our understanding of the transition it is instructive to
analyze the data for 1D systems.

We have considered 1D systems with the nearest and next-nearest
neighbor interaction\cite{tobe} and the system with LR interaction.
In the latter case we study Hamiltonian Eq.~(\ref{ham}) at filling
factor $\nu=1/2$ and $V(i-j)=1/|i-j|$.  In 1D we switch from the
homogeneous background to the chain with $\pm1/2$ charges for the
empty and occupied sites respectively.

Fig.~\ref{fig6} shows the results for the flux sensitivity vs. $J$ for
different system sizes $L$.  The sharp exponential behavior indicates
that the system becomes an insulator at small $J$.  This result
clearly contradicts to the statement by Poilblanc et al.~\cite{dag1}
that 1D Coulomb system is metallic at all $J$.

An extrapolation to $1/L\rightarrow\infty$ shown in the inset gives a
rather wide interval for $J_c$ of the IM transition between 0.17 and
0.3.

Fig.~\ref{fig7} shows few lowest eigenvalues for each quantized value
of $P$ for a cluster of 28 sites with 14 particles.  Note that the
spectrum has nontrivial symmetry around the points $P=\pm\pi/2$.  This
symmetry appears for even $N$ at $\nu=1/2$ as a result of the
particle-hole symmetry.

For even $N$ the WC icon generates two states with quasimomenta
$P=\pm\pi/2$, which are degenerate at all $J$.  As one can see from
Fig.~\ref{fig7}, at $J=0.05$ these states are separated by a gap from
the continuum of states, generated by the icon of the point defect.
At $J=0.1$ the defect band broadens and, as a result, the gap
decreases.  However, the lowest eigenvalue at $P=\pm\pi/2$ is still
separated from the defect band, whereas the second eigenvalue belongs
to it.  At this point an avoided crossing starts to develop and the
width of the gap remains almost unchanged from $J=0.1$ to $J=0.2$.  In
the latter case, the lowest eigenvalue is no longer a separated point,
but rather can be ascribed to the band.  At $J=0.3$ it becomes quite
clear that the lowest eigenvalue belongs to the continuum spectrum.
Finally, the picture at $J=1$ is almost a picture for free fermions
with the Fermi momentum $p_F=\pi/2$ and with the lowest branch
$E_{min}(P)$ close to $J|\cos(P)|$.

The proposed mechanism of the transition implies that critical value
of $J$ is determined by the energy $\Delta$ of the lowest defect at
$J=0$.  Our 2D results are summarized in the Table 1.  It shows for
comparison the middle point $J_m$ of the critical interval $\delta J$
and the zero-$J$ gap $\Delta$ for all cases we have studied.  One can
see that both $J_m$ and $\Delta$ changes by a factor of 10 depending
on the filling factor and the type of the interaction potential.

However their ratio $J_m/\Delta$ is almost constant and is close to
0.5 in all cases.  Since we assume that $J_m\rightarrow J_c$ with
increasing cluster size, this implies an empirical rule for $J_c$:

\begin{equation}
J_c=\beta\Delta_\infty
\label{cond}
\end{equation}
where $\beta$ is some number which is close to 0.5, and
$\Delta_\infty$ is the smallest energy necessary to create a point
defect in an infinitely large system.

\subsection{Study of the size effect.}

\subsubsection{ Classical size effect}

As can be expected from Table 1, the energy $\Delta$ of the lowest
excited state at $J=0$ may have much influence on $J_m$.  We show here
that $\Delta$ may have a strong size dependence in small clusters.
This kind of a size effect can be called ``classical.''

The SR and LR potentials are very different in this aspect.  In the
case of SR interaction the defect with the lowest energy is the point
defect (see Fig.~\ref{fig1}b).  The weak dependence of $\Delta$ on the
cluster size is only due to the interaction of the defect with its
images, which appear as a result of the periodic boundary conditions.

In the case of LR interaction the energy $\Delta$ depends strongly on
the size of the cluster for relatively small clusters.  This
dependence becomes stronger for smaller filling factors.  One can see
in Fig.~\ref{fig1}a that in the cluster $4\times6$ at $\nu=1/2$ the
point defect appears only as the fifth icon.  At $\nu=1/6$ the five
lowest energy icons shown in Fig.~\ref{fig1}c do not contain a point
defect at all.

We have studied thoroughly the low-energy spectrum for LR interaction
at $J=0$.  The square clusters with different sizes $L$ and filling
factors 1/2, 1/3, 1/4, and 1/6 were analyzed using classical
Monte-Carlo technique.  The results are presented in Fig.~\ref{fig8}.
At $\nu=1/3$ and 1/6 new low-energy types of dislocations appear with
increasing the cluster size.  These dislocations are restricted by the
periodic conditions in smaller clusters.  As a result, $\Delta$
decreases with size for small clusters.  However, for large enough
clusters new dislocations seize to appear, so that $\Delta$ does not
decrease.  Since the energy of a dislocation is proportional to the
size of the cluster, the point defect should win the competition in
large enough clusters.

For $\nu=1/2$ and 1/3 the point defect becomes the lowest excited
state starting with the sizes $6\times6$ and $9\times9$ respectively.
For $\nu=1/4$ and $1/6$ we are unable to find this critical size.
However, the increase of $\Delta$ with $L$ assures that the point
defect should eventually become the lowest excited state.

Our conclusion is that in the case of LR interaction one should expect
a significant size effect in $J_m$ due to the classical size effect in
$\Delta$.

\subsubsection{Quantum size effect}

Since the classical size effect is negligible for the SR interaction
we can analyze the ``quantum'' contribution to the size effect in
$J_m$ comparing the results for different clusters.  The size
dependence of $J_m$ for the SR potential can be estimated from
Fig.~\ref{fig3}b.

We study only the clusters with the dimensions commensurate with the
primitive vectors of the WC.  Otherwise the periodic continuation
destroys the crystalline order.  For $\nu=1/2$ this requires that both
$L_x$ and $L_y$ are even.  Since we can study clusters up to 16
particles this condition restricts our options to clusters $4\times4$,
$4\times6$, and $4\times8$.

The low-energy spectrum for the cluster $4\times4$ is shown in
Fig.~\ref{fig5}.  In this case the WC icon generates quasimomenta
${\bf P}=(0,0)$ and $(\pi,\pi)$.  The flux sensitivity and the ground
state splitting are shown in Fig.~\ref{fig3}b for all three clusters
studied.  One can see that the data do not show any pronounced
systematic size dependence of $J_m$, suggesting that for the SR
potential $J_c$ is within the interval 0.015---0.025.

Thus, we have found that the size effect at a given value of $\Delta$
is small.  Assuming this result to be independent of the type of
potential, one can suggest that the ``classical'' contribution is the
major for the LR interaction.  Then one can use Eq.~(\ref{cond}) to
estimate $J_c$ for the LR potential using the classical energy
$\Delta_{\infty}$.  Say, for $\nu=1/2$ we get $\Delta_{\infty}=0.61$
(see Fig.~\ref{fig8}) resulting in $J_c\approx 0.3$.  To get a
reliable estimate for this case from the quantum computations one
should consider at least $6\times6$ cluster since the point defect
becomes the lowest excited state starting with this cluster size.
\subsection{Gap at non-zero $J$}

Now we analyze the gap between the split ground state and the excited
states which belong to the defect band.  This gap is clearly seen in
Figs.~\ref{fig2}a,b, \ref{fig4}a, and \ref{fig5}.  At large $J$ the
branches have a form of beams with different slopes.  These slopes
definitely come from the confinement quantization of free fermions.
 
The large number of states in each beam reflects high degeneracy of
the free-fermion ground state at $\nu=1/2$.  Say, in Fig.~\ref{fig5}
all lines which are horizontal at large $J$ are the states that are
degenerate for the free fermions.  The splitting of these states is a
result of interaction.  The gap between the split ground state and the
bunch of the states in the same beam can be easily calculated in
mesoscopic region of large $J$, where $4\pi^2J/L^2\gg1/L$.  The
picture of beams is valid in the same region and it does not imply the
existence of a gap at large $J$ in a macroscopic system.

On the other hand, the gap $\Delta$ at $J=0$ is the energy of defect
and it has a non-zero limit in macroscopic system.  Thus, an important
question arises, whether or not the gap has a non-zero limit right
after the IM transition.  The non-zero gap would mean that the state
after the transition is superconducting.

We have made a lot of computational efforts to answer this question
but the results are still inconclusive.  Our best achievement is shown
in Fig.~\ref{fig5} where we compare the results for $4\times4$ and
$4\times8$ clusters.  The confinement quantization would prescribe
that the gap decreases in half.  We have found that the gap for the
$4\times8$ cluster is less than for $4\times4$ cluster but the ratio
is significantly larger than 0.5.

\section{Conclusions}

We have performed a numerical study of the structural and IM phase
transitions in 2D fermionic systems with Hamiltonian Eq.~(\ref{ham}).
The structural transition has been detected by studying the splitting
of the ground state, degenerate in the crystalline phase.
Simultaneously we studied the IM transition by computing the
sensitivity of the ground-state energy to the boundary conditions.  In
2D case we have studied the systems with LR and SR interactions at
different filling factors.  Within the accuracy determined by the size
effect the IM transition occurs simultaneously with the structural
transition.

We argue that the structural transition on a lattice is not of the
first order in all cases considered.  We think that the origin of the
transition is an avoided crossing of the ground state and the defect
states in the Wigner crystal with the same total quasimomentum.  This
simple picture implies that the critical value of $J$ is determined by
the defect with the lowest energy $\Delta$ at $J=0$.  To illustrate
our point the data for 1D system with Coulomb interaction are also
presented.  The possibility of the delocalized phase above the
transition to be superconducting is discussed.

We have found out that the size effect is not very strong for $J_c$ in
the case of the SR interaction.  For the LR interaction it is strong
because of the size dependence of the defect energy $\Delta$.  We
argue that a reliable estimate for $J_c$ from finite-cluster
computations in this case can be obtained with the use of the
empirical rule Eq.~(\ref{cond}).

We are grateful to A. P. Levanyuk and E. I. Rashba for helpful
discussions.  We acknowledge support of UCSB, subcontract KK3017 of
QUEST.

\begin{table}
\caption{}
\begin{tabular}{lcccc}
System & $\delta J$ & \ \ $J_m$\ \  & $\Delta$ & \ \ $J_m/\Delta$\ \ \\
\tableline
$\nu=1/2$, LR, $4\times6$:\ \ \ & 0.15---0.25   & 0.2  & 0.444  & 0.45\\
$\nu=1/2$, SR, $4\times6$:\ \ \ & 0.015---0.025 & 0.02 & 0.0448 & 0.44\\
$\nu=1/6$, LR, $6\times6$:\ \ \ & 0.01---0.03   & 0.02 & 0.037  & 0.54\\
\end{tabular}
\end{table}

\begin{figure}
\caption{Five icons with the lowest energies for (a) $\nu=1/2$, LR
interaction, (b) $\nu=1/2$, SR interaction, and (c) $\nu=1/6$ LR
interaction.
\label{fig1}}
\end{figure}

\begin{figure}
\caption{Low-energy part of the spectrum as a function of $J$ for the
cluster $4\times6$ at $\nu=1/2$ for LR (a) and SR (b) interactions.
The numbers $(n_x,n_y)$ denote the components of quasimomentum ${\bf
P}=(2\pi n_x/L_x,2\pi n_y/L_y)$.  The ground-state energy is taken as
a reference point.
\label{fig2}}
\end{figure}

\begin{figure}
\caption{Flux sensitivity for two directions of vector potential and
the ground-state splitting as a function of $J$ at $\nu=1/2$.  (a) LR
interaction for a cluster $4\times6$. (b) SR interaction for clusters
$4\times4$, $4\times6$, and $4\times8$.  Dashed lines show large-$J$
and small-$J$ asymptotic as obtained by the best fit with correct
powers of $J$.  The vertical bars show the critical region of the
transition.
\label{fig3}}
\end{figure}

\begin{figure}
\caption{Results for $\nu=1/6$, cluster $6\times6$.  (a) Low-energy
part of the spectrum; (b) flux sensitivity for different $P$ and two
directions of vector potential.  The numbers $(n_x,n_y)$ denote the
values of {\bf P} as in Fig.~2.  The reference point is taken to be
$A+BJ^2$, where $A$ is the energy of the WC at $J=0$, and $B=177$ is
the exact $J^2$ term as obtained from perturbation theory.  Dashed
lines have the same meaning as in Fig.~3.  The vertical bars show the
critical region of the transition.
\label{fig4}}
\end{figure}

\begin{figure}
\caption{The same as Fig.~2b for the clusters $4\times4$ and
$4\times8$.  The data for $4\times8$ cluster is presented for ${\bf
P}=(0,0)$ (dots) and ${\bf P}=(\pi,\pi)$ (circles) only.
\label{fig5}}
\end{figure}

\begin{figure}
\caption{Flux sensitivity in units $J/L$ as a function of $J$ for the
1D system with LR Coulomb interaction at $\nu=1/2$ for different sizes
$L$.  Long-dashed line shows the theoretical value $L\delta E/J=\pi$
for free fermions.  The inset shows the extrapolation to
$1/L\rightarrow0$.
\label{fig6}}
\end{figure}

\begin{figure}
\caption{The dependence of the total energy on the total quasimomentum
at different $J$ for 1D system with LR Coulomb interaction at
$\nu=1/2$ and size $L=28$.
\label{fig7}}
\end{figure}

\begin{figure}
\caption{Size dependence of the lowest excitation energy at $J=0$ for
different filling factors as obtained by classical Monte-Carlo.  The
saturation occurs at such size when the point defect becomes the
lowest excitation.
\label{fig8}}
\end{figure}

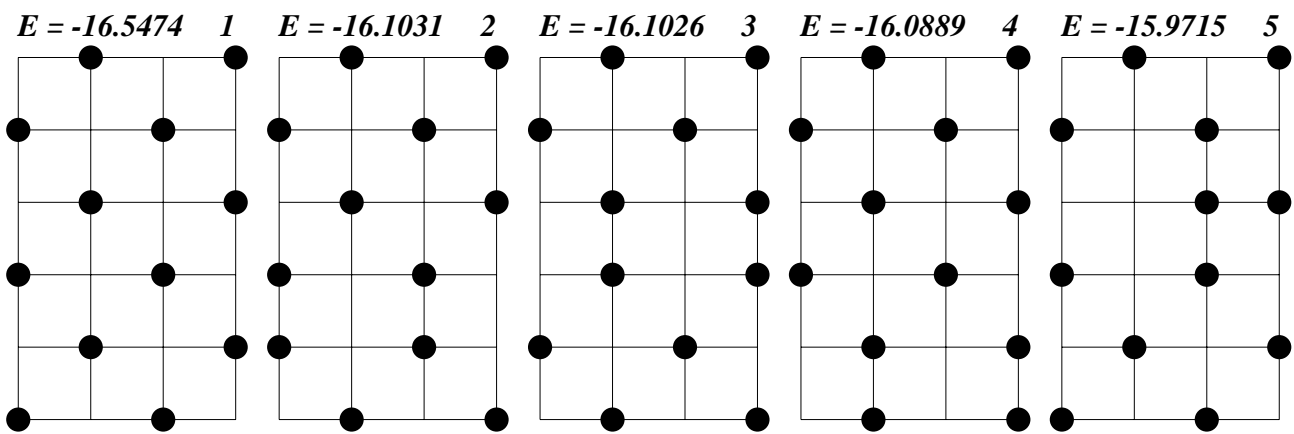
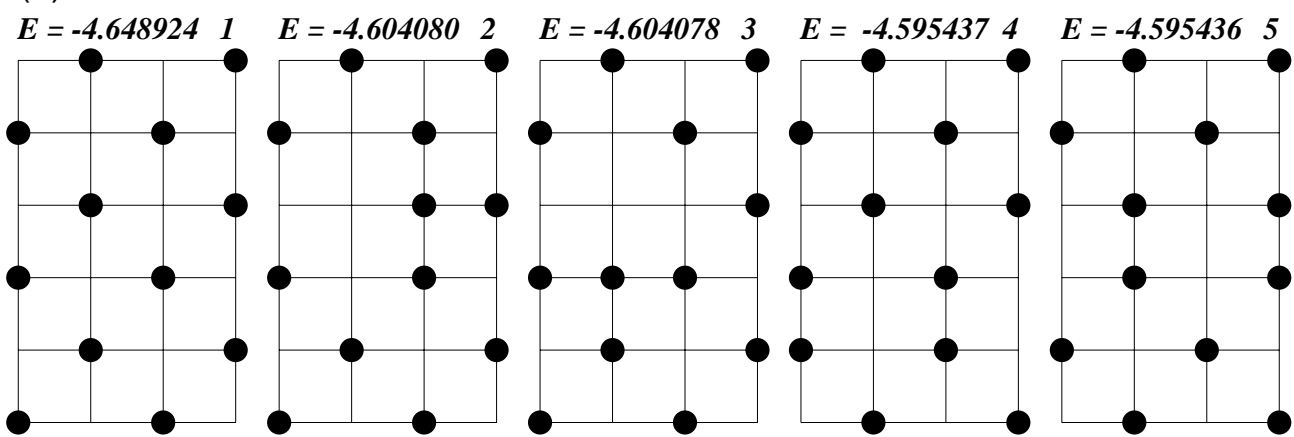
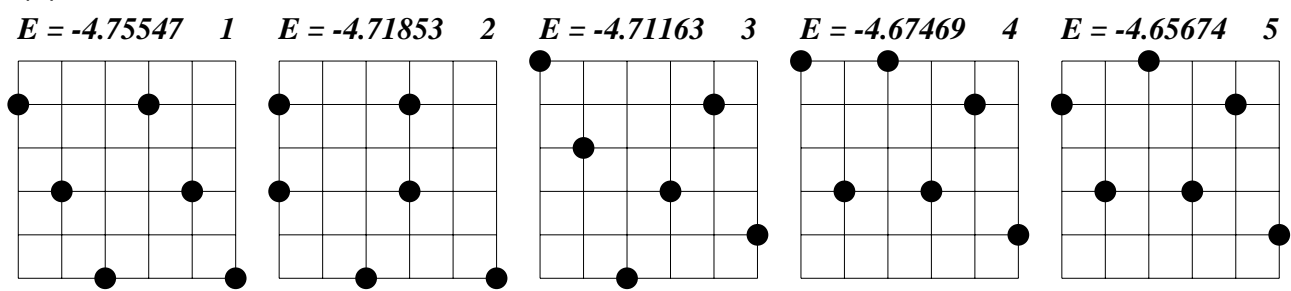
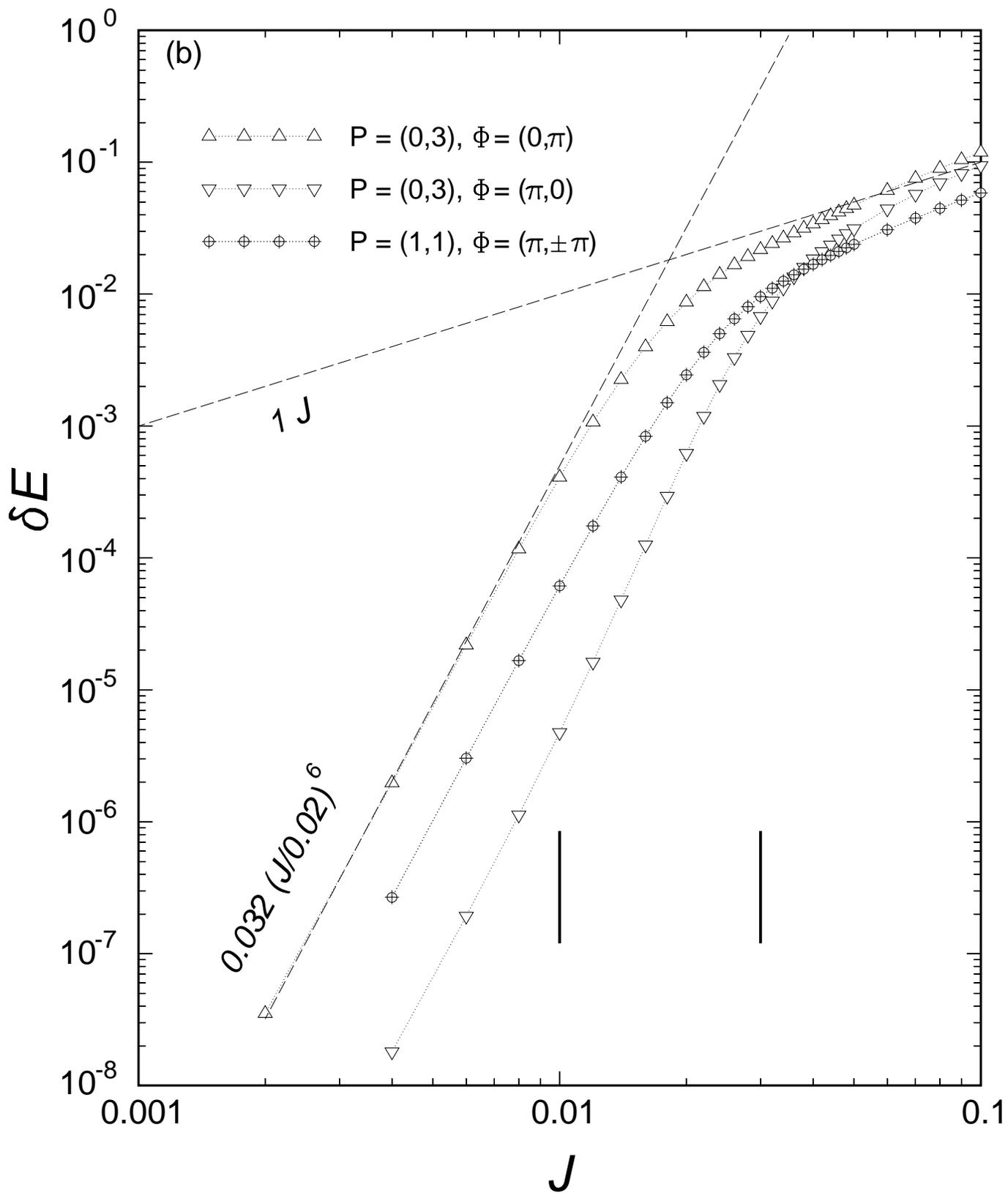
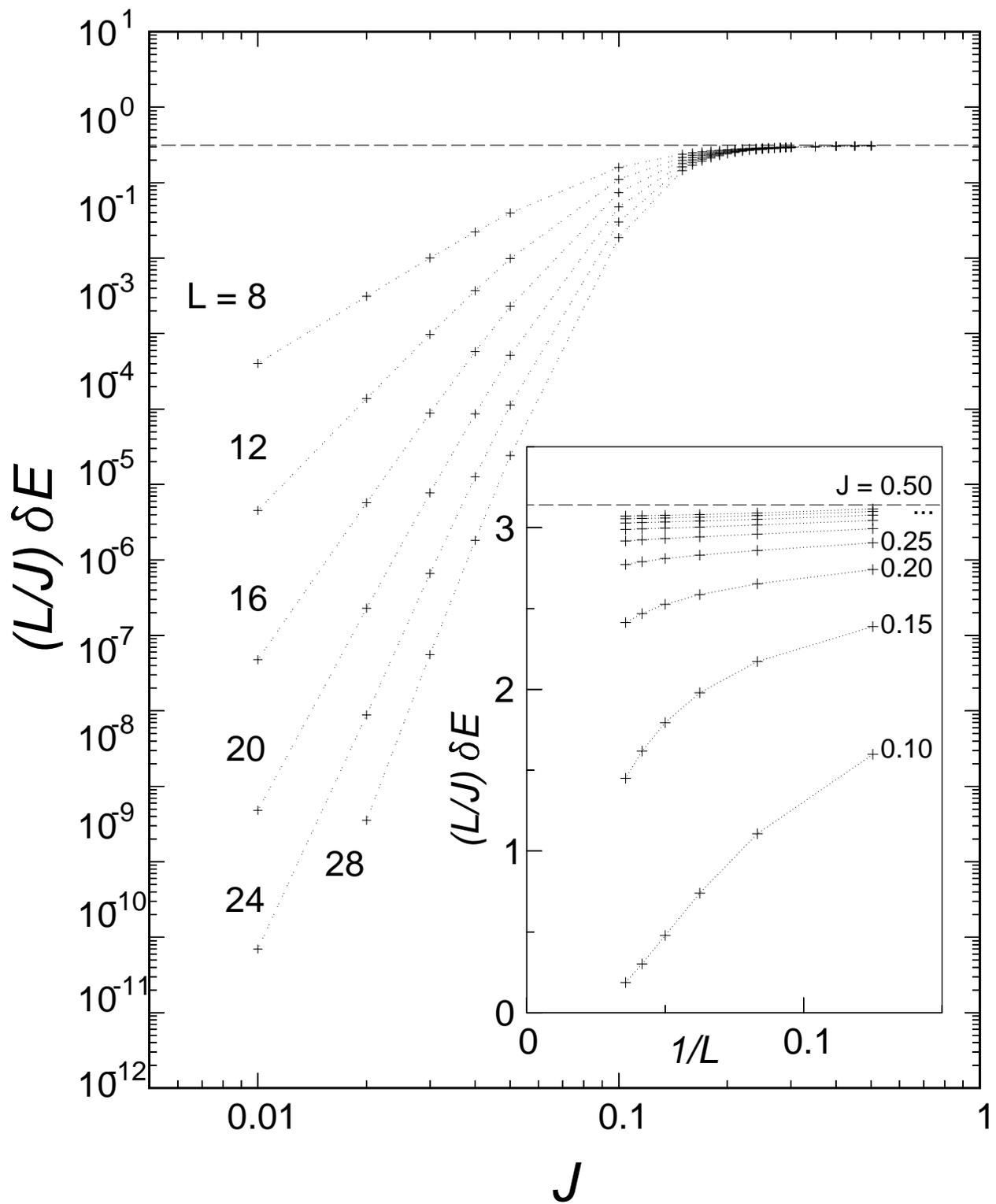
\end{document}